\documentclass[3p,twocolumn]{elsarticle}

\usepackage{graphicx}
\usepackage{amssymb}
\usepackage{bm}

\begin{document}

\begin{frontmatter}
\title{Quantum Oscillation Studies of the Fermi Surface of LaFePO}
\author[a]{A. Carrington}
\author[a]{A. I. Coldea}
\author[a]{J. D.~Fletcher}
\author[a]{N. E. Hussey}
\author[a]{C. M. J. Andrew}
\author[a]{A. F. Bangura}
\author[b]{J. G. Analytis}
\author[b]{J.-H. Chu}
\author[b]{A. S. Erickson}
\author[b]{I. R. Fisher}
\author[c]{R. D. McDonald}
\address[a]{H.~H.~Wills Physics Laboratory,  University of Bristol, Bristol, BS8 1TL, UK.}
\address[b]{Geballe Laboratory for Advanced Materials and Department of Applied Physics,\\ Stanford University,
Stanford, California 94305-4045, USA.}
\address[c]{National High Magnetic Field Laboratory, Los Alamos National Laboratory, \\ MS E536, Los Alamos, New
Mexico 87545, USA.}

\begin{abstract}
We review recent experimental measurements of the Fermi surface of the iron-pnictide superconductor LaFePO using
quantum oscillation techniques. These studies show that the Fermi surface topology is close to that predicted by first
principles density functional theory calculations, consisting of quasi-two-dimensional electron-like and hole-like
sheets.  The total volume of the two hole sheets is almost equal to that of the two electron sheets, and the hole and
electron Fermi surface sheets are close to a nesting condition. No evidence for the predicted three dimensional pocket
arising from the Fe $d_{z^2}$ band is found.  Measurements of the effective mass suggest a renormalisation of around
two, close to the value for the overall band renormalisation found in recent angle resolved photoemission measurements.
\end{abstract}

\end{frontmatter}

\section{Introduction}

Experimental determinations of the Fermi surface are important for a number of reasons.  Firstly, although calculations
of the band structure using density functional theory within the local density approximation (and its variants) give a
good general picture, in many cases the calculated band positions are found to be accurate to only a few tens of meV
\cite{CarringtonMCBHYLYTKK03,MackenzieJDMRLMNF96,CarringtonY07}. These relatively small errors can lead to substantial
differences in the size and shape of the Fermi surface. For the iron-pnictides this will have a bearing on, for
example, how close to perfect nesting the various Fermi sheets are and also the magnitude and anisotropy of the Fermi
velocities. This will in turn lead to corrections to calculations of the magnetic excitation spectrum and the
interactions that may be responsible for superconductivity, plus other physical properties, such as $H_{c2}$ and its
anisotropy. Secondly, conventional band structure calculations only include many-body effects at the mean field level,
and so in strongly interacting systems there can be substantial renormalisation of the effective masses or Fermi
velocities from electron-phonon or electron-electron interactions. These mass renormalisations need to be determined
experimentally. Thirdly, in systems with structural or magnetic instabilities (such as charge density waves and spin
density waves) there may be subtle changes in the lattice structure which could lead to a reconstruction of the Fermi
surface, possibly breaking up a large Fermi surface into many small pockets.

Magnetic field induced quantum oscillations effects in, for example, the magnetisation  [the de Haas-van Alphen effect
(dHvA)] or the resistance [Shubnikov-de Haas effect (SdH)] have been used for more than half a century to determine the
Fermi surfaces of metals \cite{Shoenberg}.  The power of the technique is that it provides very accurate measurements
of the Fermi surface cross-sectional areas (typically with an accuracy better than 0.1\% of the area of the Brillouin
zone), effective masses (typically with an accuracy better than $\sim 5$\%), and quasiparticle scattering rates.  There
are, however, several experimental constraints.  The most serious is that the dHvA signal is damped exponentially by
impurity scattering, which broadens the Landau levels and therefore reduces the size of the anomaly which occurs when
they cross the Fermi level.  This reduction is given by the Dingle factor, $R_D=\exp(-\pi k_F/eB\ell)$, where $k_F$ is
the orbitally averaged Fermi wavevector for the particular Fermi surface orbit, and $\ell$ is the inelastic
quasiparticle mean-free-path (which is often shorter than the corresponding quantity derived from transport properties
as low angle scattering events are not weighted out). This means that dHvA oscillations are only observable at high
magnetic field in very pure samples. Finite temperature also reduces the size of the oscillatory signal due to the
thermal smearing of the discontinuity in occupation number at the Fermi level. The signal is reduced by a
factor 2 when $T \simeq 0.12\,B/m^*$ ($m^*$ is the quasiparticle effective mass measured in units of the free electron
mass $m_e$). For an effective mass of $\sim 2$ and magnetic field of 15~T the amplitude of the signal is reduced by
a factor ten from its zero temperature value at $T \simeq 2$~K, with a further attenuation of approximately one order
of magnitude per Kelvin at higher temperature. Performing experiments at such low temperatures means that for most
pnictide materials we are usually in the regime where the upper critical field is very high ($>$60T). Although it is
possible to observe dHvA oscillations in the superconducting mixed state,
\cite{JanssenHHMSW98,FletcherCKK04,Graebner76} the signal suffers further attenuation as the field is reduced below
$H_{c2}$, \cite{Maki91} hence oscillations are typically only observed near to $H_{c2}$.

A complication with understanding dHvA results is that it is not possible to determine experimentally the $k$-space
location of the observed Fermi surface orbits, and so interpretation of the data usually relies on a comparison with
band-structure calculations. Despite these limitations, quantum oscillations are the most precise technique to
determine Fermi surface geometry, including full information about the three dimensional topology.

Another technique which has been used extensively in recent years to determine Fermi surfaces is angle resolved
photoemission (ARPES). The ARPES technique has moderate resolution of the in-plane momentum but can provide a direct
$k$-resolved map of the Fermi surface \cite{DamascelliRMP2003}. The $k_z$ momentum cannot be probed directly although
the $k_z$ point at which the in-plane momentum is measured may be controlled to some extent by varying the photon
energy. Unlike dHvA which is a bulk probe, ARPES probes just a few unit cells below the surface. The termination of the
surface can be an issue, especially if the compound does not cleave on an electrically neutral layer \cite{Hossain08}.
ARPES and dHvA are therefore complementary techniques, which both have their strengths and weaknesses.

At the time of writing, dHvA experiments have been reported in just two iron-pnictide materials. Superconducting LaFePO
\cite{ColdeaFCABCEFHM08,SugawaraSDMKYO08}, which will be the focus of this article and undoped SrFe$_2$As$_2$
\cite{SebastianGHLSML08}. This latter compound is non-superconducting and has a antiferromagnetic ground state - which
makes the Fermi surface reconstruct into many small pockets.

The iron phosphide superconductor LaFePO, is isostructural with superconducting LaFeAsO$_{1-x}$, and the two materials
are predicted to have very similar electronic structure.  Its relatively low $T_c \sim$ 6~K \cite{KamiharaHHKYKH06} has
been linked to the fact that the Fe-P bond angles depart substantially from those of a regular tetrahedron
\cite{CHLee08}. There is some controversy about whether stoichiometric LaFePO is superconducting, or whether a few
percent of oxygen vacancies are present even in the nominally undoped material \cite{Analytis2008,McqueenRWHLHWGC08}.
Oxygen deficiencies of around 1\% are difficult to determine on small single crystals and so for the moment this
remains an unresolved question.  One significant difference from the arsenide compounds is that LaFePO is not magnetic
even in its non-superconducting state \cite{McqueenRWHLHWGC08} and so we do not expect any Fermi surface
reconstruction. Another difference is that recent penetration depth measurements\cite{fletcher09} suggest a
superconducting order parameter with line nodes, rather than the fully gapped state that has been observed in several
As-based materials \cite{Hashimoto08a,Malone08}. This may mean that the order parameter symmetry in iron-pnictides may
 vary between materials.
 A detailed knowledge of the differences in electronic structure may be important in understanding how this arises.

The low $T_c$ of LaFePO leads to a correspondingly low value of $\mu_0H_{c2}\simeq 0.7$\,T, for $H\|c$.  This, combined
with the very high purity levels found for flux grown crystals (residual resistance ratios up to $\sim$ 90) makes
LaFePO an excellent choice for dHvA studies. To date, two groups have reported dHvA results for LaFePO.  This article
will focus mainly on our own data \cite{ColdeaFCABCEFHM08} but will include comparisons to the data of Sugawara
\textit{et al.} \cite{SugawaraSDMKYO08} where appropriate.

\section{Experimental details}
\subsection{Sample growth and characterization} Single crystals of LaFePO were grown from a molten Sn flux.  La,
Fe$_2$O$_3$, P and Sn mixed in the molar ratio 3:1:2:24, were placed in alumina crucibles and sealed in quartz tubes
under a small partial pressure of argon. Typically the melt was held at 1190$^\circ$ C  for 18hrs, before cooling at
10$^\circ$C/hour to 650$^\circ$C. The remaining liquid was decanted at the base temperature using a centrifuge, and the
crystals removed from the crucible. Excess Sn on the surface of the crystals was removed by etching in dilute HCl
followed by an immediate rinse in methanol.  Measurements of the magnetic penetration depth \cite{fletcher09} showed no
evidence for any Sn flux remaining on the samples (with a resolution of about 1 $\mu$m$^3$).  Crystals of up to 0.4\,mm
on a side were produced by this process although, as all our dHvA measurements were performed using the
micro-cantilever method, only small single crystals were selected.  It is likely that the smaller crystals would be
more homogeneous than the larger ones. Some growth runs were made by an alternative route using La$_2$O$_3$ instead of
Fe$_2$O$_3$ to introduce oxygen into the melt.  Details are described in Ref.\ \cite{Analytis2008}. Sugawara \textit{et
al.} \cite{SugawaraSDMKYO08} used a very similar Fe$_2$O$_3$ based sample preparation route.

\begin{figure}
\centering
\includegraphics[width=7cm]{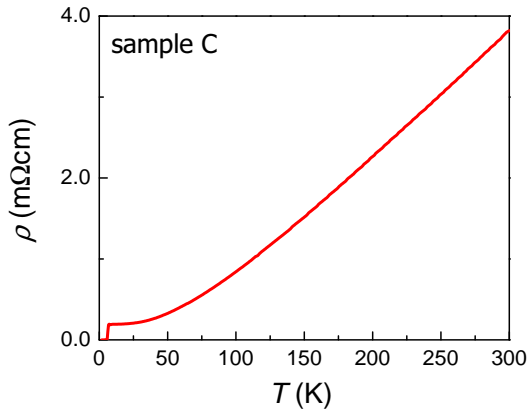}
\caption{The temperature dependence of the \textit{interlayer} resistivity of LaFePO (sample C).} \label{figrho}
\end{figure}

\begin{figure}
\centering
\includegraphics[width=7cm]{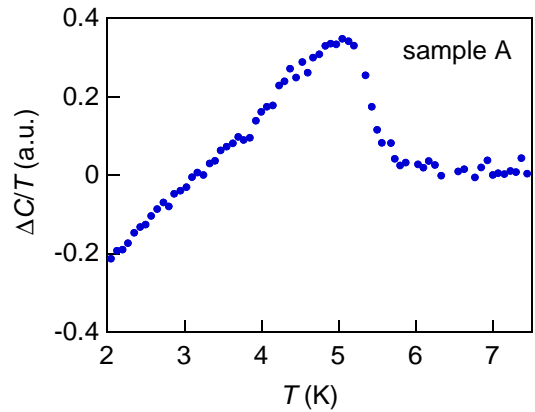}
\caption{Specific heat anomaly at the superconducting transition in a single crystalline sample of LaFePO (sample A).
$\Delta C=C(0)-C(B)$, i.e., the zero field specific heat with normal state background subtracted by using data taken in
a field large enough to suppress superconductivity ($B=1$\,T).} \label{figcp}
\end{figure}

The crystals were characterized by x-ray diffraction, electrical resistivity and specific heat.  The x-ray results show
that crystals are tetragonal, space group Pnmm, with cell dimensions $a$ = 3.941(2)\AA, $c$ = 8.507(5)\AA~and atomic
positions $Z_{\rm P}$=0.63477 and $Z_{\rm La}$=0.1489 in agreement with other previous results \cite{KamiharaHHKYKH06}.
The crystallographic orientation of the crystals was identified using Laue diffraction.

Resistivity measurements were conducted using a standard low frequency ac, 4-probe technique, with contacts applied
with Epotek H20E paste. Fig.\ \ref{figrho} shows the $c$-axis resistivity of sample $C$, which indicates conventional
metallic behaviour, superconductivity at low temperature, and a residual resistance ratio $R(300 K)/R(T_c) \sim 52$.
Similar resistance data are also reported by Sugawara \textit{et al.} \cite{SugawaraSDMKYO08}.   In-plane resistivity
has also been measured on other samples and the anisotropy $\rho_c/\rho_a$, which was relatively independent of
temperature, was in the range 13 to 17 \cite{Analytis2008}.

Heat capacity was measured using an ac technique, with modulated light as a heating source and a 12$\mu$m thermocouple
as a thermometer.  This allowed us to measure the same small single crystals that were used for the dHvA study (typical
dimensions $100\times 100 \times 20 \mu$m$^3$). The data shown in Fig.\ \ref{figcp} shows a sharp, well defined anomaly
at the superconducting transition ($T_c$ midpoint = 5.8\,K) showing that there is bulk superconductivity in our LaFePO
samples.

\subsection{Micro-cantilever torque measurements}

Due to the small size of the best crystals of LaFePO, the most suitable techniques for measuring quantum oscillations
are micro-cantilever torque or resistivity.  For the torque measurements the samples were attached with quick dry epoxy
to the end of the self-sensing microcantilever (Seiko instruments model: SSI-SS-ML-PRC120).  The resistance of the
cantilever is measured with a conventional Wheatstone bridge arrangement, using an ac current excitation of typically
$I=20\mu$A at a frequency of 72\,Hz.  The cantilever was mounted on a single axis rotator and placed in a $^3$He
cryostat, in the bore of a 19\,T superconducting magnet (in Bristol) or the 45\,T hybrid magnet in Tallahassee,
Florida.  We have used this technique previously for studies of MgB$_2$ \cite{YellandCCHMLYT02} and cuprate
superconductors \cite{BanguraFCLNVHDLTAPH08,VignolleCCFMJVPH08}.

Sugawara \textit{et al.} \cite{SugawaraSDMKYO08} used the same cantilevers, but with a field modulation technique; a dc
current is passed through the cantilever and an additional oscillatory field ($B_{ac}=10$\,mT, $f=$11\,Hz) is applied
yielding voltage oscillations directly related to the torque.

\section{dHvA results}
\begin{figure}
\centering
\includegraphics[width=8cm]{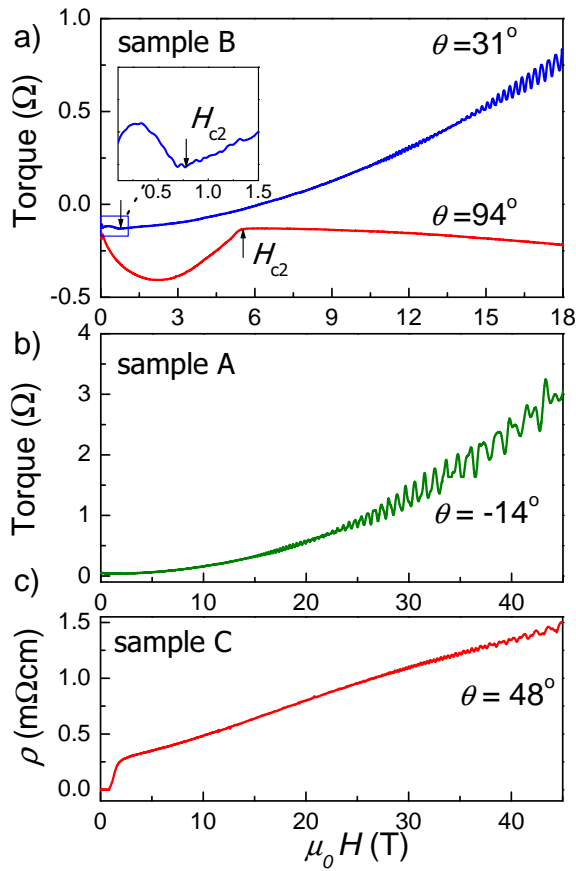}
\caption{ a) Torque measurements on LaFePO measured at $T = 0.35$~K and in magnetic fields up to 18~T for different
magnetic field directions (sample B). The arrow indicates the position of $H_{\rm c2}$. b) Torque measurements up to
45~T for sample A. c) Field dependence of the interlayer resistivity in magnetic fields up to 45~T for sample C.}
\label{fig_raw_data}
\end{figure}
Fig.~\ref{fig_raw_data} shows raw torque and resistivity measurements in high magnetic fields up to 45~T and at low
temperatures $T\approx 0.35(5)$~K on different samples of LaFePO. In low magnetic fields the torque signal is
characteristic of a type II superconductor in the vortex state with $\mu_0 H_{c2}^{\|c} \approx 0.68$~T
(Fig.~\ref{fig_raw_data}a). As the system is rotated away from $H || c$ the position of  $H_{c2}$ shifts to much higher
values ($\mu_0 H_{c2}^{\perp c} \approx 7.2$~T). These are the characteristics of an anisotropic superconductor with an
anisotropy of $\gamma_{\xi} \approx 10$ \cite{anisotopynote}, similar to that found in SmFeAsO$_{1-x}$F$_x$
\cite{Daghero} but much larger than that for (Ba,K)Fe$_2$As$_2$ \cite{Yuan08}. The torque response in LaFePO is mostly
reversible for field sweeps up and down suggesting weak vortex pinning in our crystals.

\begin{figure}
\centering
\includegraphics[width=8cm]{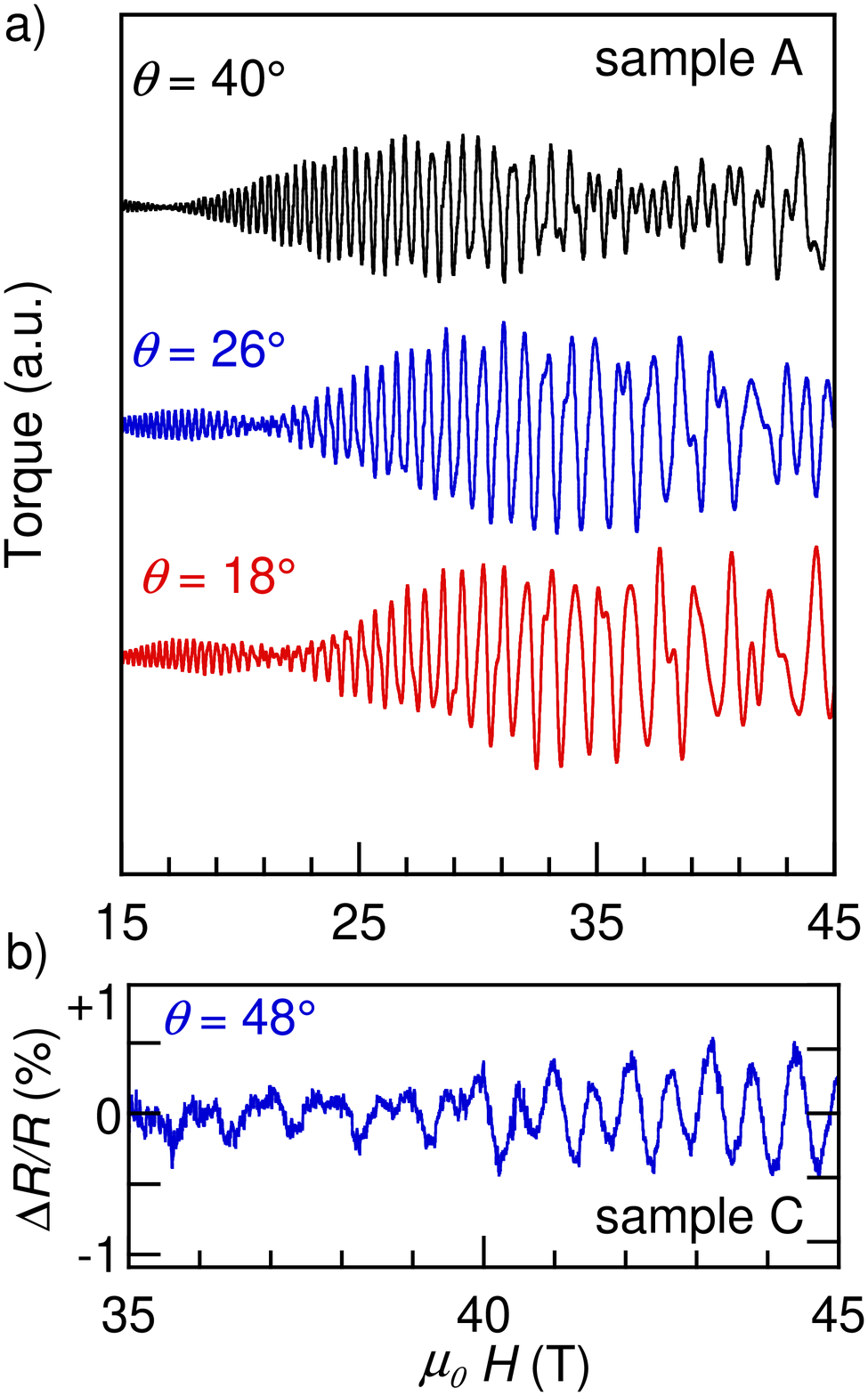}
\caption{a) Oscillatory part of the torque (dHvA) (sample A) and b) resistance (SdH) (sample C) in magnetic fields up
to 45~T for different magnetic field angle, $\theta$, with respect to the $c$ axis.} \label{fig_subtracted_raw_data}
\end{figure}

For magnetic fields higher than $H_{c2}$ the background torque signal varies quadratically with $H$ as expected for a
simple paramagnet. Above 9~T the quantum oscillations start to become visible superimposed on the background signal.
The interlayer resistance shows a positive magnetoresistance and above 30~T a weak oscillatory signal is visible
(Fig.~\ref{fig_raw_data}c). By subtracting the background signal from the torque and resistivity data using a
 polynomial (5$^{\rm th}$ order) the richness of quantum oscillation signal is clearly visible (Fig.\
\ref{fig_subtracted_raw_data}).

\begin{figure}
\centering
\includegraphics[width=7cm]{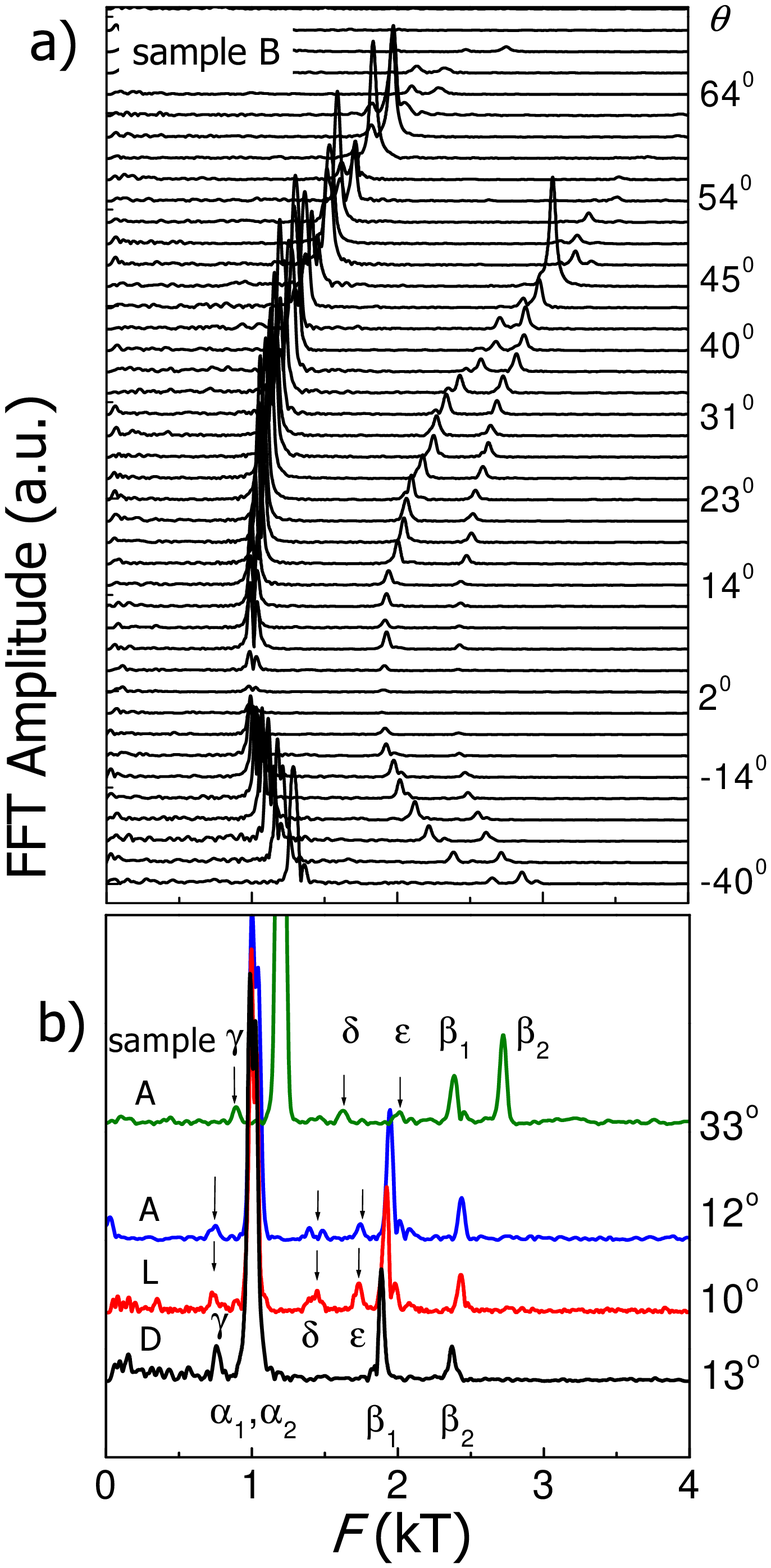}
\caption{ a) The angle dependence of the Fourier transform spectra on LaFePO measured at $T = 0.35$~K for the field
range 10-18~T (sample B). The large peak near 45$^{\circ}$ is due to the enhancement of the dHvA at the 'Yamaji angle'
(see text). Different runs are shifted for clarity and the angle step corresponds to $\approx 3^{\circ}$. b) Fourier
transform spectra for different field orientations for sample A (field range 15-45~T), sample L (field range 15-50~T)
and sample D (field range 10-18~T). The positions of the orbits with smaller amplitudes, $\gamma$, $\delta$, and
$\varepsilon$ are shown by arrows. The positions of the main frequencies corresponding to orbits near extremal areas of
Fermi surface of LaFePO are shown by their labels $\alpha_1$, $\alpha_2$, $\beta_1$ and $\beta_2$.}
\label{fig_rotationplot}
\end{figure}

The fast Fourier transform (FFT) of the oscillatory data for different samples is shown in Fig.\
\ref{fig_rotationplot}. We  identify the dHvA frequencies $F$, which are related to the extremal cross-sectional areas
$A_{k}$, of the FS orbits via the Onsager relation, $F = (\hbar/2\pi e) A_{k}$ as the peak in the FFT. The angular
dependence of these frequencies as the magnetic field angle, $\theta$, is rotated away the $c$ axis ($\theta =
0^\circ$) towards the conducting planes ($\theta = 90^\circ$) \cite{angle}allows us to construct a three-dimensional picture of the
shape and size of the Fermi surface. Fig.\ \ref{fig_rotationplot}a shows such a detailed angular
dependence for sample B, which was measured up to 18~T at $T=$0.3(1)~K. In this particular experiment four different
oscillatory components can be seen, whose frequency and amplitude is strongly dependent on angle. The frequencies shift
to higher values as $\theta$ is increased. This is described approximately by $F \propto 1/\cos \theta$, strongly
suggesting that the signals arise from orbits on quasi-two-dimensional Fermi surfaces sheets. The amplitude of the
oscillatory part of torque,
\begin{equation}
\tau_{osc} \propto -\frac{1}{F} \frac{dF}{d\theta} M_{\parallel} B \label{eqtorque}
\end{equation}
(see Ref.\ \cite{Shoenberg}), is determined by the Fermi surface anisotropy and thus vanishes near symmetry axes and
for spherically symmetric Fermi surfaces.

\begin{figure*}
\centering
\includegraphics[width=16cm]{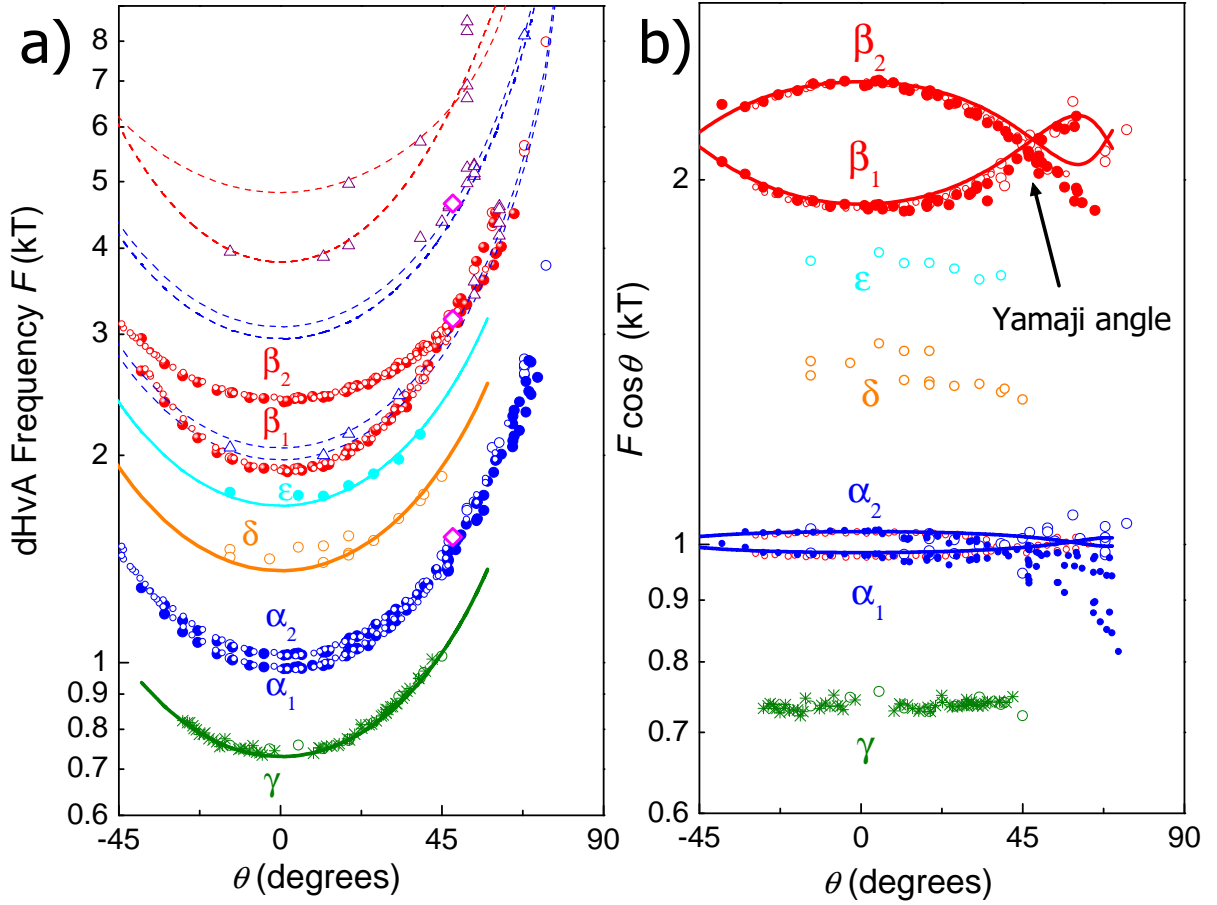}
\caption{ a) Angle dependence of all observed frequencies. Different symbols correspond to: sample A (open circles),
sample B (filled circles), sample C (diamonds) and sample D (stars). Possible harmonics of the main frequencies are
shown by triangles and the dashed lines indicate their calculated location. Solid lines overlapping the $\gamma$,
$\delta$ and $\varepsilon$ frequencies are fits to a 1/cos $\theta$ dependence. b) Angle dependence of $F \cos\theta$.
Solid lines are calculations for a simple cosine warped cylinder (see text). The position of the `Yamaji angle' is
indicated by the arrow.  Note the logarithmic vertical axis on both parts of the
figure.}\label{fig_yamaji}\end{figure*}

\begin{table*}
\centering \caption{Experimental effective masses ($m^*$) and frequencies from dHvA data for samples A, B and D (our
results) and sample S (after Ref.\cite{SugawaraSDMKYO08}).}

\begin{tabular}{c   c  c c c  c   c c c l}
\hline \hline
  &&\multicolumn{3}{c}{$F$(kT)} &&\multicolumn{3}{c}{$m^*/m_e$} \\
%&  & $F$(kT)&        &&  &$m^*/m_e$ &  &\\
\hline
Branch      &~~~~~& (A, B)  &       D &  S    &~~~~~&  A &   B    & S & \\
    \hline
$\alpha_1$  &&   0.985(7)   &0.962    &0.963 &&  1.9(2)  &  1.8(1) &   2.13(4)&\\
$\alpha_2$  &&   1.025(7)    &1.002   &0.998  && 1.9(2)  &  1.8(1) &   2.34(7)&\\
$\beta_1$   &&   1.91(1)     &1.824   &1.88   && 1.7(2)   &  1.8(1)  &  1.82(16)& \\
$\beta_2$   &&   2.41(1)     &2.331   &2.38   && 1.8(2)  &  1.9(1) &   2.28(16) &\\
$\delta$    &&   1.36(2)     &        &        &&  1.8(3)  &         & &\\
$\gamma$    &&   0.73(2)     &0.734   &        && 1.7(3)  &          &&\\
$\epsilon$  &&   1.69(2)     &       &        &&  2.1(3)  &          &&\\
\hline
\end{tabular}
\label{table_freqs}
\end{table*}

Four frequencies have significantly larger amplitudes than the others. The first two of these, which we label
$\alpha_1$ and $\alpha_2$, have very similar frequencies, with values near $F \approx 1$~kT ($\Delta F \approx 35$~T).
Near H$\|$c these frequencies are separated enough for distinct peaks to be observed directly in the FFT. At higher
angles these peaks tend to merge, but beating can be seen in the raw data, indicating the survival of two components,
whose frequencies can be found by fitting the torque data directly. Two higher frequencies, $\beta_1$ ($\approx
1.9$~kT) and $\beta_2$ ($\approx 2.4$~kT), also have a rather strong amplitude and have frequencies that rise rapidly
as $\theta$ is increased, crossing near $\theta = 45^\circ$.

Frequencies $\alpha_{1},\alpha_2, \beta_1$ and $\beta_2$ were observed in all measured samples, A, B, D and L, as shown
in Fig.~\ref{fig_rotationplot}. Sugawara {\em et al.} \cite{SugawaraSDMKYO08}, also observed these four orbits with
almost identical frequencies and angle dependence.

In further experiments additional frequencies were observed. Sample D was studied in fields up to 18~T, the same range
as sample B, but comes from the batch with a higher residual resistivity ratio ($\sim 85$). In this crystal we observe
an additional orbit, denoted $\gamma$, with $F\approx 0.7$~kT which has a similar amplitude to the $\beta_2$ orbit.
Samples A and L were measured in much higher fields (up to 50~T) and there we observed in addition to $\alpha$, $\beta$
and $\gamma$ a further two distinct frequencies which we label $\delta$ and $\varepsilon$. These were not observed by
Sugawara {\em et al.} \cite{SugawaraSDMKYO08}, probably because of the lower maximum field available in their
experiment. The values of all the observed frequencies, extrapolated to H$\|$c, are listed in Table \ref{table_freqs}.
We observe a very good agreement for all the measured samples, with the frequencies agreeing to within a few percent.

The angular dependence of the frequencies as the magnetic field is tilted away from the $c$ axis is shown in
Fig.~\ref{fig_yamaji}. For a two-dimensional cylindrical Fermi surface the dHvA frequency varies like $1/\cos{\theta}$
and deviations from this indicate the degree of warping. This is shown in Fig.~\ref{fig_yamaji}b by plotting $F(\theta)
\cos \theta$. The Fermi surfaces of branches $\alpha_{1}$, $\alpha_{2}$ and $\gamma$ are close to perfect cylinders as
well as those for $\delta$ and $\varepsilon$ (within the limitation of the available data).  The upward curvature on
the $F\cos\theta$ plot of the $\beta_1$ data shows that this orbit originates from a section of Fermi surface that is
locally concave along the $c$-axis (e.g., the minimum of a cosine warped tube), and likewise $\beta_2$ is from a convex
(maximum) section.

For a quasi-two-dimensional Fermi surface with weak cosine $c$-axis warping, Yamaji \cite{Yamaji89} has shown that the
angular dependence of the two frequencies, $F_{\pm}$, arising from extremal areas of a single warped cylinder is given
by,
\[
F_{\pm}(\theta) \cos\theta = (\hbar/2\pi e)[\pi k_F^2 \pm 4 \pi m_e t_c J_0(c k_F \tan\theta)].
\]
In this expression, the first term is the mean frequency $(\hbar/2\pi e)\pi k_F^2=(F_{\rm min}^0+F_{\rm max}^0)/2$, the
prefactor of the second term $(\hbar/2\pi e)4 \pi m_e t_c$ is $(F_{\rm max}^0-F_{\rm min}^0)/2$ and $J_{0}$ is the
Bessel function ($c$ is the lattice parameter and $t_c$ parameterises the warping). Here $F_{\rm min}^0$ and $F_{\rm
max}^0$ represent minimum and maximum frequencies at $\theta=0$, and may be taken directly from the data, so this
equation has no free parameters.  A plot of this function is shown in Fig.~\ref{fig_yamaji}b, and provides a very good
description of the data.  For certain angles, $\theta$, where the Bessel function $J_0=0$ the two frequencies cross. At
this point all the cross section areas are equal and their contributions to the total magnetization interfere
constructively to give a sharp maximum in the amplitude of the signal (peak effect). This `Yamaji angle' is observed
close to $\theta \sim 45(2)^{\circ}$ for the $\beta$ orbits in LaFePO, (see Fig.~\ref{fig_yamaji}b) and the peak effect
is clearly seen in Fig.~\ref{fig_rotationplot}a. These observations show convincingly that $\beta_1$ and $\beta_2$ are
the minimum and maximum frequencies of the same quasi-two-dimensional Fermi surface sheet.  A similar behaviour, albeit
with much smaller warping, is observed for the $\alpha$ frequencies indicating that these too originate from a single
quasi-two-dimensional Fermi surface sheet.

\subsection{Torque interaction}

The total torque response, both background and oscillatory, causes a deflection of the lever, and so the angle $\theta$
is not quite constant during the field sweep. This effect produces harmonics of the main frequencies (indicated in
Fig.~\ref{fig_rotationplot}a) and may also produce mixing of the main frequencies, such as
$\beta_{1,2}\pm\alpha_{1,2}$, $2\alpha$, $2\beta$, etc. and so it is important to identify any such frequencies. Close
to $\theta=0$ we find that the frequency of $\delta$ ($F_\delta$) is very close to $F_{\beta_2}-F_{\alpha_2}$. However,
if $\delta$ was produced by torque interaction we would expect it to have an angle dependence intermediate between
$\beta_2$ and $\alpha_2$. This does not appear to be the case (see Fig.\ \ref{fig_rotationplot}b), however more
detailed data will be required to settle the issue. The $\gamma$ and $\varepsilon$ frequencies cannot be generated by
this effect.

\begin{figure}
\centering
\includegraphics[width=7cm]{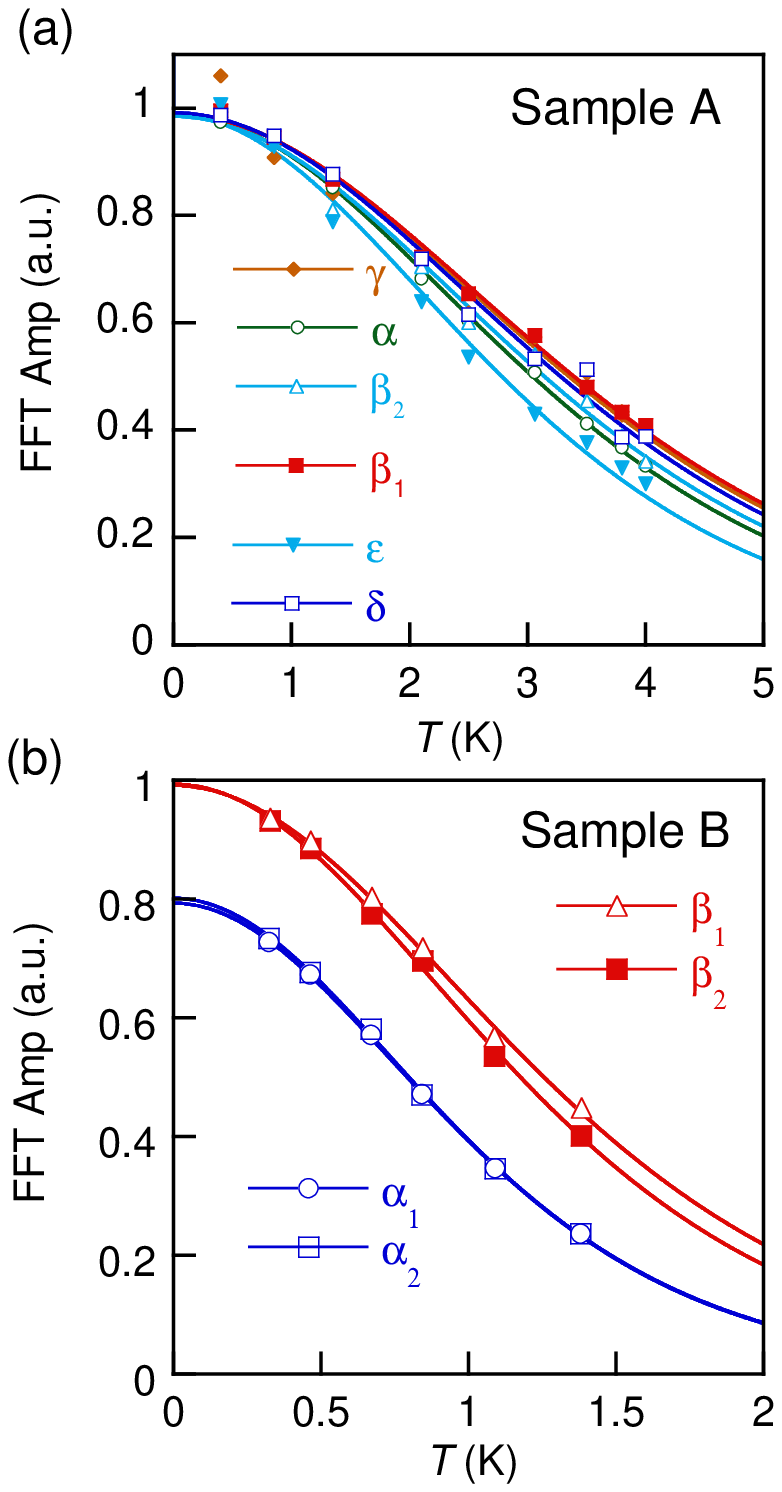}
\caption{The temperature dependence of the Fourier amplitude for sample A ($\theta = 2^\circ$ and field window between
25~T and 45~T). The inset shows the temperature variation of the normalized amplitudes for the main frequencies for
sample B ($\theta = 32^\circ$) (data are offset for clarity). Solid lines are fits to extract the effective mass (see
text)} \label{fig_meff}
\end{figure}

\subsection{Effective mass}

The cyclotron effective mass of the quasiparticles on the various orbits were determined by fitting the temperature
dependent amplitude of the oscillations to $R_T=X/\sinh(X)$ where $X=14.69m^*T/B$ \cite{Shoenberg}, as shown in
Fig.~\ref{fig_meff}. These measurements were made with the field close to the $c$-axis and the value for
$\theta=0^{\circ}$ was inferred using the usual approximation that the mass scales with the dHvA frequency (for sample
B the mass was measured at several different angles to confirm this). The obtained masses range between 1.7~{\it m$_e$}
and 2.1{\it m$_e$} and are listed in Table~\ref{table_freqs}. These values are slightly smaller than the those reported
in Ref.\ \cite{SugawaraSDMKYO08}, but within the quoted errors.

From the field dependence of the amplitude at constant temperature we can estimate the inelastic quasiparticle
mean-free-path, $\ell$, from the Dingle factor $R_D$. We find a mean free path of $\approx 1300$~\AA~ and 800~\AA~ for
$\alpha$ and $\beta$ orbits, respectively. A similar value of 940~\AA~ was found for the $\beta_1$ orbit by Sugawara
\emph{et al.}\cite{SugawaraSDMKYO08}. For the $\delta$, $\gamma$ and $\varepsilon$ orbits the scattering rate is a
factor $\sim 2-3$ larger than that for the $\alpha$ orbits although the small size of the signal precludes any detailed
analysis.

\section{Band structure}

\begin{figure*}
\centering
\includegraphics[width=16cm]{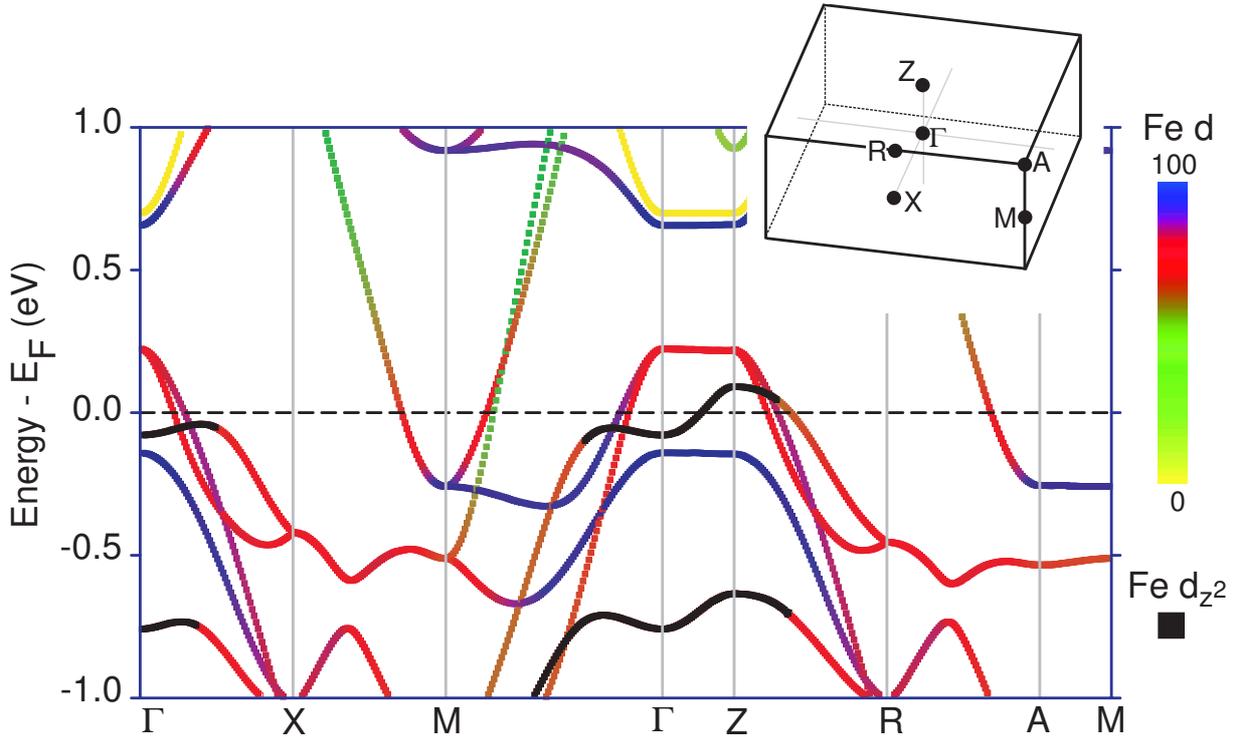}
\caption{The calculated band-structure of LaFePO (without spin-orbit corrections) along principle symmetry directions.
The colour scale reflects the total Fe $d$ character. Bands with Fe $d_{z^2}$ character greater than 50\% are coloured
black.  The inset shows the Brillouin zone with symmetry labels.} \label{spagplot}
\end{figure*}

\begin{figure}
\centering
\includegraphics[width=6cm]{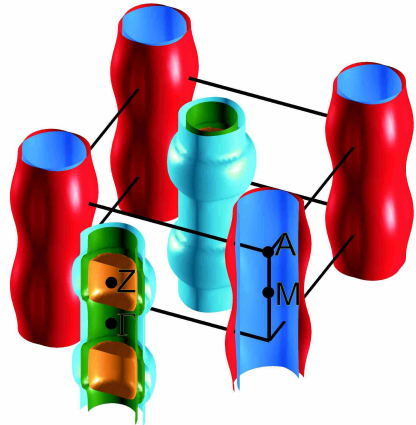}
\caption{The calculated Fermi surface of LaFePO (without spin-orbit corrections).} \label{fsfigure}
\end{figure}

The band structure of LaFePO has been calculated by Leb\`{e}gue \cite{Lebegue07}, and consists mainly of tubular two
dimensional sheets of Fermi surface; two hole like tubes at the zone centre and two electron-like tubes at the zone
corner.  In addition there is a more three dimensional hole-like pocket near the Z point (see Fig.~\ref{fsfigure}).

We have repeated these calculations so that we can calculate the dHvA parameters and make a detailed comparison to our
data. Our own calculations were carried out using the WIEN2K package \cite{wien2k}, which is an implementation of a
full-potential, augmented-plane-wave plus local-orbital scheme. We used a generalized-gradient approximation form for
the exchange correlation potential \cite{PerdewBE96} and the experimental crystal structure. We used $10^4$ $k$-points
in the full Brillouin zone for convergence and generally $10^5$ $k$-points for the Fermi surface and dHvA parameter
calculations. Larger $k$-point densities were used for the two-dimensional cuts (typically $10^5$ $k$-points in each 2D
cut).  The calculations were performed with and without the spin-orbit correction which was found to produce small but
important differences in the band structure.

\begin{figure}
\centering
\includegraphics[width=6cm]{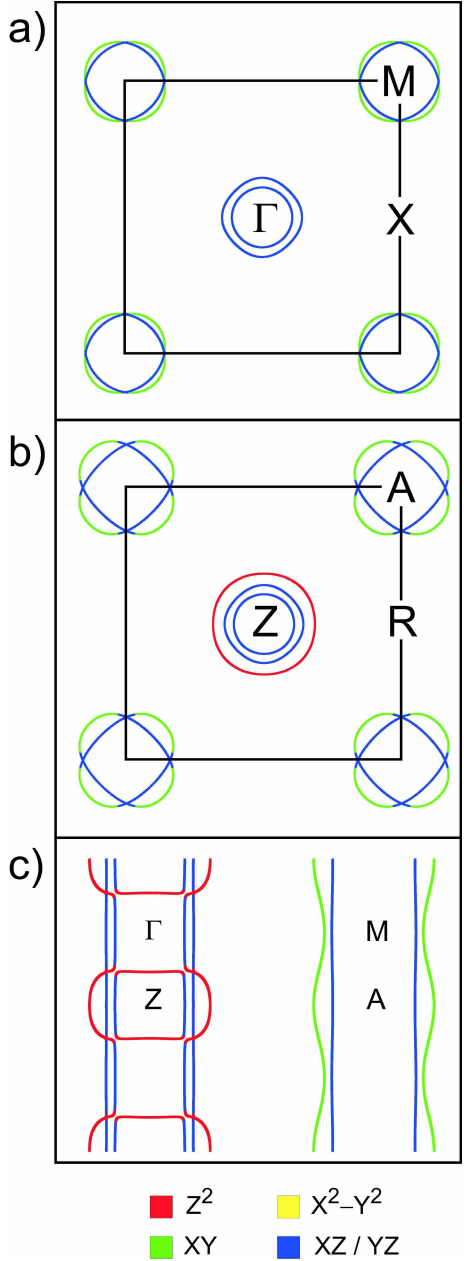}
\caption{Slices through the Fermi surface calculated without spin-orbit interaction. The slice are (a) along (001) at
the centre of the zone, (b) along (001) at the top of the zone and (c) along (110) through the centres of both electron
and hole sheet. The symmetry points are indicated. Each $k$ point is coloured according to the maximum band character at
that point (see key at bottom of figure).} \label{nospinorbitSlices}
\end{figure}

The band structure without spin-orbit corrections is shown in Fig.\ \ref{spagplot}. These results are in good agreement
with those of Leb\`{e}gue \cite{Lebegue07}. The bands crossing the Fermi level are mostly of Fe $d$ character as
emphasized in the plot, and mostly weakly dispersive along the $c$-direction (see the sections of the plot $\Gamma
\rightarrow$ Z, and A$\rightarrow$ M). The exception is a band which has mostly Fe $d_{z^2}$ character which is
emphasized in black in the plot.  In Fig.\ \ref{fsfigure} we show the resulting Fermi surface and in Fig.\
\ref{nospinorbitSlices} two dimensional slices of the Fermi surface emphasizing the character of the bands.  In Fig.\
\ref{nospinorbitSlices} the plot is colour coded according to the strongest band character at each individual $k$-point.
At the zone corner there are two electron-like quasi-two-dimensional Fermi surface sheets, which touch along the
$\langle 100\rangle$ directions.  In the Brillouin zone corresponding the iron sublattice (which is $\sqrt{2}$ larger
and rotated by 45$^\circ$ compared to the actual Brillouin zone corresponding to the real crystal structure
\cite{IIMazin08032740}), these sheets would become two distinct ellipses. They fold onto each other in the actual
Brillouin zone. The inner electron sheet has mostly $d_{xz}$ ($d_{yz}$) character whereas the outer one has mostly
$d_{xy}$ character. The inner sheet has significantly less $c$-axis warping than the outer sheet.  Close to the zone
center are three hole-like sheets. There are two tubes running along the $c$-axis, both sheets have strong $d_{xz}$
($d_{yz}$) character although the outer one has almost equal $d_{x^2-y^2}$ character. These tubes are cut through by
the $d_{z^2}$ band which forms a squashed spherical shape. Hence, we are left with two tubular sheets with strong
warping near $Z$ and a third three dimensional closed cylinder pocket. The position of this $d_{z^2}$ band is strongly
dependent on the phosphorous atom position ($Z_{\rm P}$) and also on the doping (which we simulated using the virtual
crystal approximation).

\subsection{Effect of spin-orbit interaction}
A significant effect of the spin-orbit interaction is that, as shown in  Fig.\ \ref{spinorbitSlices}, it splits the two
electron Fermi surface sheets into two, with a gap in $k$-space equal to $\sim$1.2\% of $2\pi/a$.  The bands are
non-longer degenerate along the X-M (R-A) line. The gap is 50meV close to the Fermi level along X-M and 40meV along
R-A.  Another effect is that it also increases the gap at the points where the $d_{z^2}$ band crosses the hole tube
sheets, and decreases the warping of these Fermi surface sheets. The changes to the Fermi surface are summarized in
Fig.\ \ref{spinorbitSlices}.

\subsection{Magnetic breakdown}

Given that the gaps are quite small there is a question as to whether the electrons may tunnel between these Fermi
surface sheets when subjected to the strong magnetic fields used in our study. This effect, which is known as magnetic
breakdown, has been studied extensively and is found to occur when the field is in excess of a `breakdown field'
$H_{BD}$. As this critical field is crossed the electrons would begin to change their orbits so that they follow the
trajectory of the ellipse which would exist if there were no gap.  There is a smooth crossover at $H_{BD}$; for $H \ll
H_{BD}$ orbits on the separate inner and outer electron sheets would be visible and for  $H \gg H_{BD}$ there would be
just one orbit on the joined up ellipse (or two if $c$-axis warping is taken into account).  In the intermediate region
both frequencies corresponding to both types of orbit would be observed. In general frequencies corresponding to
$F_\alpha+\frac{1}{4}n(F_\beta-F_\alpha)$, where $n$=1,2 or 3 might be expected \cite{hayden}.  An estimate of the
breakdown field can be made using the following formula \cite{Chambers}
\[
\mu_0H_{BD}= \frac{\pi\hbar}{2e}\left(\frac{k_g^3}{a+b}\right)^\frac{1}{2}
\]
 where $k_g$ is the $k$-space gap between the two orbits (measured at the closest point) and $a^{-1}$ and $b^{-1}$ are the $k$-space
 radii of curvatures of the orbits on each side of the gap.  Using the parameters from our calculations we estimate $\mu_0H_{BD} \simeq 80$\,T.
 This would increase by a factor $\sim$3 to if the real experimentally determined gaps are used (see later).  As
 $F_\delta\simeq F_{\alpha_2}+(F_{\beta_2}-F_{\alpha_2})/4$ and $F_\varepsilon \simeq
 F_{\alpha_2}+(F_{\beta_2}-F_{\alpha_2})/2$ these two frequencies could arise from magnetic breakdown, although our
 estimate of  $H_{BD}$ is a factor $2-6$ higher than our maximum field.  To rule this out completely measurements on higher quality
  samples are required, so that we see if these orbits can be observed at much lower field.

We note here that although the Fermi surfaces will be Zeeman split by the strong applied field, the measured dHvA
frequencies actually correspond to to the Fermi surface area $A_k(H)-H\partial A_k(H)/\partial H$ (i.e., the area at
a particular field extrapolated to zero field with the local slope $\partial A_k/\partial H$ )\cite{Ruitenbeek82}.  So
only non-linear splitting (which we do not expect for LaFePO) gives rise to extra dHvA frequencies in high field.

\subsection{Comparison to data}

\begin{figure*}
\centering
\includegraphics[width=16cm]{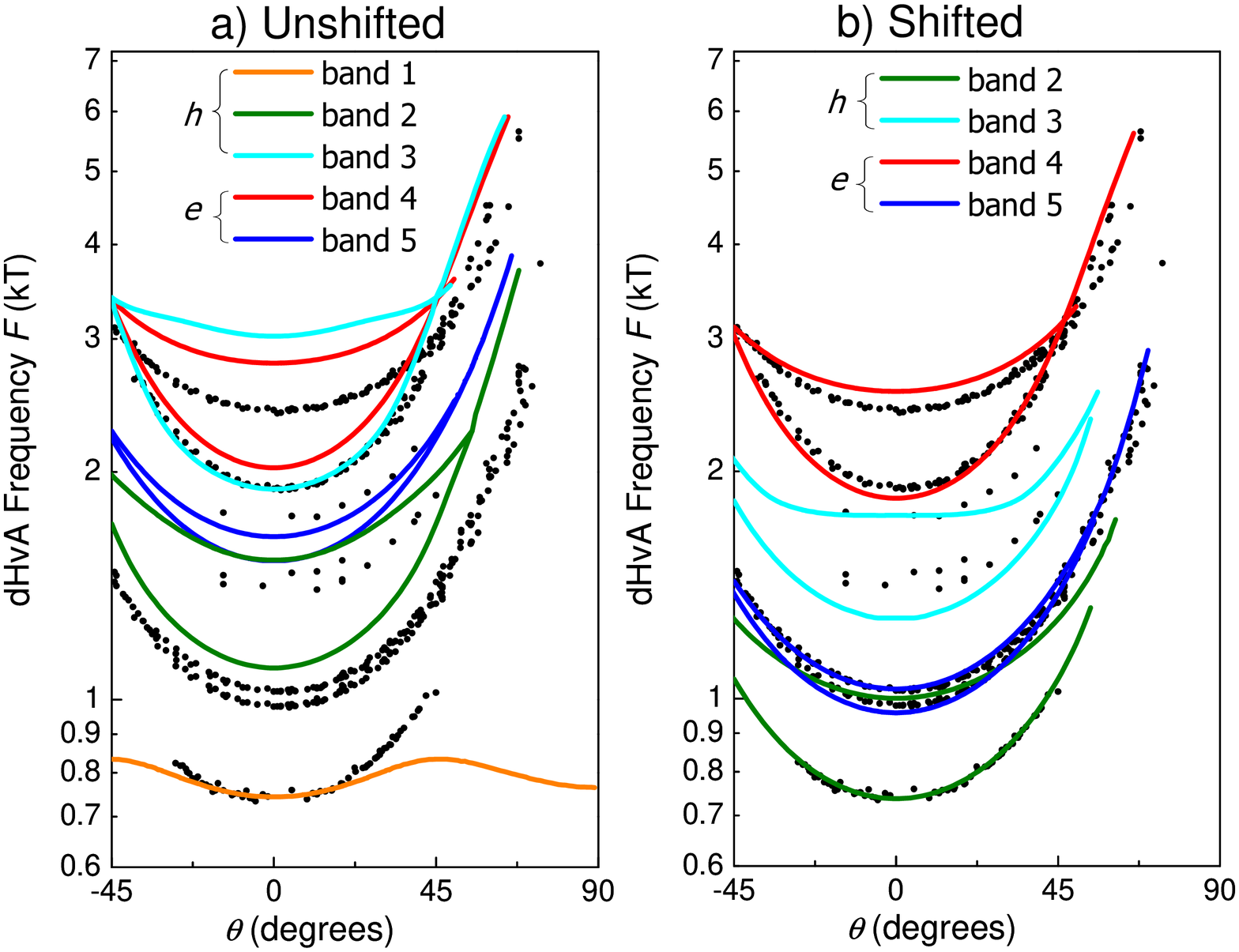}
\caption{Calculated dHvA frequencies versus field angle.  The experimental data are shown by black dots. a) shows the
calculation (including spin-orbit interaction) without any band shifting , and b) with the bands shifted as described
in the text.} \label{CalcdHva}
\end{figure*}

To compare with the experimental results we calculate the extremal cross-sectional areas of each Fermi surface sheet
and plot these as a function of magnetic field angle as the field is rotated from parallel to $c$ to parallel to $a$.
The results shown in Fig.\ \ref{CalcdHva}, and numerical values are given in Table \ref{table_calcs}.  Note that
rotations from $c$ towards (110) (i.e., at 45$^\circ$ to the $a$-axis) yield results which are almost indistinguishable
on the scale of Fig.\ \ref{CalcdHva} (the differences are only clearly apparent as the field approaches the $a-b$
plane). For each tubular sheet of Fermi surface there are two branches corresponding to the maximal and minimal
frequencies. The difference in minimum value and angle dependence reflects the degree of warping of the tube.

\begin{figure*}
\centering
\includegraphics[width=16cm]{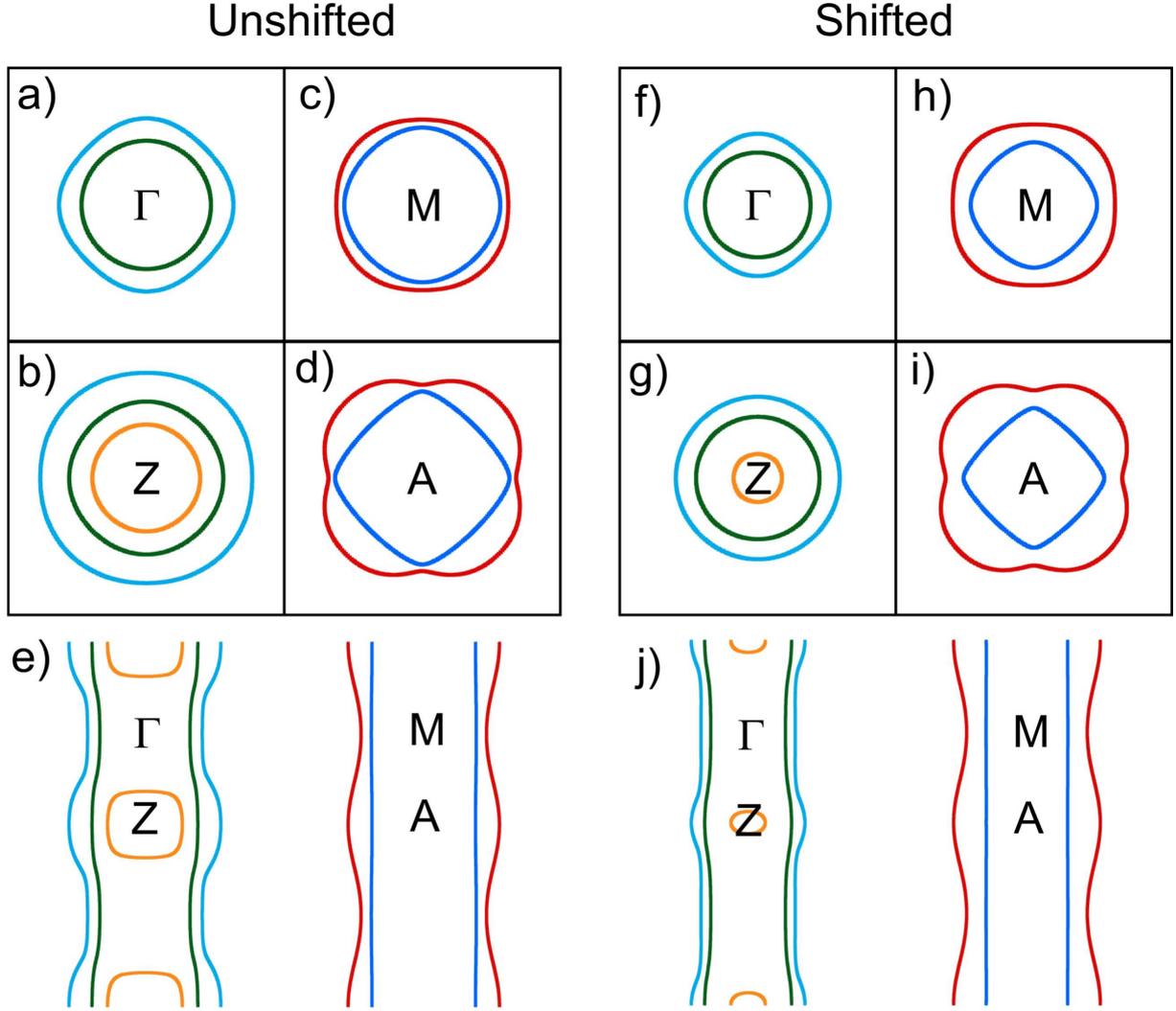}
\caption{Fermi surface slices including the spin-orbit interaction (compare to Fig.\ \ref{nospinorbitSlices}) (a-d) are
(001) slices showing (a,b) the hole sheets at $\Gamma$,Z and (c,d) the electron sheets at M,A, (e) shows a (110) slice.
The right panels show the same cuts but the with bands shifted to best fit the dHvA data (as described in the text)}
\label{spinorbitSlices}
\end{figure*}

\begin{table*}
\centering \caption{Calculated dHvA frequencies and band masses $m_b$. Results are shown with and without the rigid
band shifts described in the text. The results of the fully relativistic band structure calculation reported in
Ref.\cite{SugawaraSDMKYO08} are also included for comparison (denoted S).}

\begin{tabular}{c c ccc c cccc}
\hline \hline
& ~~~~&\multicolumn{3}{c}{$F$(kT)}&~~~~~&\multicolumn{3}{c}{$m_b/m_e$}\\
\hline
Branch         && unshifted  &  shifted   & S  & &   unshifted  &   shifted    &  S &\\
 \hline
1Z-$h$         &&   0.741    &   0.125  &-&        &1.0 & 1.9 & -&\\
2$\Gamma$-$h$  &&   1.100    &   0.741  &0.708&    &0.9 &0.7& 0.547&\\
2Z-$h$         &&   1.527    &   0.992  &0.781&    &1.2 &1.1&0.689&\\
3$\Gamma$-$h$  &&   1.898    &   1.276  &1.050&    &1.8 & 1.1&0.771&\\
3Z-$h$         &&   3.026    &   1.721  &1.170&    &2.9 & 2.5&0.868&\\
4M-$e$         &&   2.023    &   1.846  &0.919&    &0.7 & 0.7&0.490&\\
4A-$e$         &&   2.787    &   2.554  &0.979&    &0.9 & 0.9&0.521&\\
5M-$e$         &&   1.527    &   0.957  &0.657&    &0.8 & 0.7&0.376&\\
5A-$e$         &&   1.641    &   1.048  &0.836&    &0.9 & 0.8&0.544&\\
\hline
\end{tabular}
\label{table_calcs}
\end{table*}

The calculation for the outer electron sheet (band 4) is in good agreement with the data for the $\beta$ frequencies,
both in absolute size and warping.   The inner electron sheet (band 5) is mostly likely associated with the $\alpha$
frequencies. The angle dependence of band 5 and its warping is in good agreement with the data but the absolute value
is $\sim$ 30\% too large. Another reason for believing this assignment is that the scattering rate (and absolute signal
amplitude) for the $\alpha$ frequencies is similar to the $\beta$ frequencies, which is understandable as both arise
from the same section of Fermi surface before the spin-orbit interaction split them apart.  All the $\alpha$ and
$\beta$ orbits have very similar effective masses which is also understandable if they originate from essentially the
same Fermi surface.  The calculations are brought into much better agreement with the data if band 4 and 5 are shifted
by +30meV and +85 meV respectively (see Fig.\ \ref{CalcdHva}(b)).  Note that this has the effect of increasing the
splitting between the two bands by a further 32\,meV (recall the spin-orbit induced splitting was $\sim$ 50meV). Hence,
with these shifted bands there is now quite sizeable gap in $k$-space between the two Fermi surface sheets (see Fig.\
\ref{spinorbitSlices}.)

Without the strongly $c$-axis dispersive $d_{z^2}$ band there would be just two hole-like cylinders at the zone centre
with very weak $c$-axis warping.  This would give rise to two very close frequencies, with 1/cos $\theta$ angle
dependence for each sheet.  Indeed, this is exactly what is found for the minimum frequencies of bands 2 and 3. The
presence of the $d_{z^2}$ band causes significant warping of each of these sheets near their maximum cross-sections
with a `bulge' around the Z point, which is roughly spherical locally (see Fig.~\ref{spinorbitSlices}e). Consequently
the angle dependence of the maximal frequencies for these orbits is very weak, and the maxima have much higher
frequencies than the minima. The small closed three-dimensional cylindrical pocket (band 1), has a roughly
$1/\cos\theta$ dependence close to $\theta=0$ and $1/\sin\theta$ near to $\theta=90^\circ$.

With the $\alpha$ and $\beta$ orbits identified, it is natural then to identify the weaker $\gamma$, $\delta$, and
$\varepsilon$ branches with the hole orbits.  Given that we do not observe experimentally any orbits close to
$\theta=90^\circ$, and the remaining orbits vary like $1/\cos\theta$ up to at least 45$^\circ$, it seems that the three
dimensional pocket (band 1) has not been observed.  The lowest frequency we observe $\gamma$ can then be identified
with the minimum of the inner hole tube. If we shift all the hole bands by -53\,meV we find almost perfect agreement
between band 2 and $\gamma$, and then the minimum of band 3 coincides with $\delta$ without any further adjustment.
With this shift the maximum of band 2 overlaps with the $\alpha$ orbits so may be difficult to observe experimentally.
The maximum of band 3 coincides with the $\varepsilon$ orbit although it does not have the correct curvature.

Given that we see no evidence for band 1, and that the observed frequencies all have a $1/\cos\theta$ angle dependence,
so that there is no evidence for the `bulge' close to the Z point (see Fig.\ \ref{spinorbitSlices}e), it is likely that
in reality the Fe $d_{z^2}$ does not actually cross the Fermi level \cite{bulgenote}.  As we noted above this band is
the most sensitive to P position and doping. So it could be that $\gamma$ corresponds to both frequencies from the now
weakly warped inner hole pocket and $\delta$ (or possibly) $\varepsilon$ corresponds to both frequencies from the now
weakly warped outer hole pocket. The problem with this is that it leaves $\varepsilon$ (or $\delta$) unidentified.  Of
course, this difficulty would be removed if $\delta$ was really caused by torque interaction or magnetic breakdown.
Further experiments, with lower scattering rates for the hole orbits, as well as following how the frequencies move
with doping and pressure will hopefully resolve these issues.

\begin{table}
\begin{center}
\caption{Calculated number of electrons ($n$) or holes in the first Brillouin zone of LaFePO for each Fermi surface
sheet, according to the band-structure calculations (with spin-orbit corrections) with ($n_S$) and without ($n_{NS}$) shifting the bands.
The negative sign on the electron count indicates holes. Note this corresponds to two formula units. The shifts
($\Delta E$) are in units of meV.}

%\begin{tabular}{|c|l|l|l|}
%\hline
%Band&$n_{NS}$&$\Delta E$&$n_S$\\
%\hline
%1&$-$0.016&  $+$53&$-$0.001\\
%2&$-$0.098&  $+$53&$-$0.062\\
%3&$-$0.180&  $+$53&$-$0.105\\
%4&$+$0.178&  $-$30&$+$0.162\\
%5&$+$0.118&  $-$85&$+$0.074\\
%\hline total&   +0.002&&+0.069\\
%\hline \end{tabular} \label{ElectronCountingTable}
%\end{center}
%\end{table}

\begin{tabular}{@{~~} c  r@{}l @{~~} r@{}l @{~} r@{}l @{~~}} \hline\hline Band &\multicolumn{2}{c}{$n_{\rm NS}$} &\multicolumn{2}{c}{$\Delta E$} &\multicolumn{2}{c}{$n_{\rm NS}$}\\ \hline
1&$-$0&.016&   $-$&53&$-$0&.001\\
2&$-$0&.098&   $-$&53&$-$0&.062\\
3&$-$0&.180&   $-$&53&$-$0&.105\\
4&$+$0&.178&   $+$&30&$+$0&.162\\
5&$+$0&.118&   $+$&85&$+$0&.074\\
\hline
total&   +0&.002&&&+0&.069\\
\hline \end{tabular} \label{ElectronCountingTable} \end{center} \end{table}

From valence considerations we expect LaFePO to have exactly equal number of electrons and holes.  In Table
\ref{ElectronCountingTable} we show calculations of the number of electron/holes in each of Fermi surface sheets for
both shifted and unshifted bands. The very slight departure of the unshifted bands total from zero (0.002 electrons per
unit cell) reflects the precision of the numerical integration and Fermi surface determination.  For the shifted bands
the imbalance is 0.067 electrons per unit cell. In principle, this could suggest that there either remains some sheet
of Fermi surface which is undetected, or our assignment of the orbits is incorrect.  A possible candidate for the
former explanation is the presence of the 3D hole pocket which we have not observed - however given the likely small
size of this it is unlikely to make up this difference. Alternatively, if we assume that the inner hole pocket is
$\gamma$ and the outer one is $\varepsilon$ (and attribute $\delta$ to the torque interaction) the imbalance would be
approximately halved, 0.036 electrons per unit cell.   This small remaining imbalance could easily be explained by a
slight oxygen deficiency in our samples.  An imbalance of 0.036 electrons per unit cell corresponds to an oxygen
deficiency of just 0.9\% which is below our x-ray detection limit.

The band structure and dHvA frequencies for LaFePO were also calculated by Sugawara {\em et al.}
\cite{SugawaraSDMKYO08} using an in-house fully-relativistic code. The Fermi surface again is composed for quasi-two
dimensional electron and hole sheets however in their results the $d_{z^2}$ does not cross the Fermi level. In this
calculation the sizes of the orbits are all much smaller than those found in our work (as well as those found
experimentally, see Table\ref{table_freqs}). For example, the frequencies for all of the electron orbits are less than
1 kT compared to 1.9-2.4\,kT found here (see Table \ref{table_calcs}).

With the various orbits provisionally assigned, we can now compare the effective masses.  The theoretical values for
both shifted and unshifted bands are shown in Table \ref{table_calcs}.  For most orbits the band masses are close to
$1$. The larger values for some of the hole orbits are caused by the top of the $d_{z^2}$ passing close to the Fermi
level. Comparing with the measured effective masses we can see that the enhancement factor is approximately 2 for all
the observed orbits.

By approximating the Fermi surface sheets as unwarped cylinders, we can estimate the correspondence between our
measured effective masses and the electronic specific heat coefficient ($\gamma_E$).  In the 2D limit, the contribution
to $\gamma_E$ from each two-dimensional sheet is given by $\frac{\pi}{6}\frac{m^*}{\hbar^2}a^2N_Ak_B^2$.  Four
quasi-two-dimensional sheets with $m^*$=2.0 $m_e$ would therefore correspond to 6 mJ/molK$^2$ (note that here a mol
refers to a formula unit not the unit cell).  Experimentally, a value of $\gamma_E=7.0\pm0.2$ mJ/mol K$^2$ was
estimated for our crystals \cite{Analytis2008} which is close to our dHvA estimate.

One remaining question about the bandstructure is the influence of the magnetic interactions.  The experimentally
determined positions of the La and P atoms are not in their lowest energy positions according to the DFT calculations.
The calculated optimal position of the P atom is around 0.05\,\AA~from its experimental position. This error is reduced
considerably if the ground state is assumed to be antiferromagnetic  \cite{mazinjohannes08} - which it is not
experimentally found to be. This problem may originate from the mean-field nature of the DFT calculations which do not
take into account the fluctuation effects which can suppress the magnetism \cite{mazinjohannes08}. Clearly there are
some internal consistency problems with the calculations and some authors have chosen to base their calculations on the
optimized atomic positions rather than the experimental ones \cite{LuYMEACSHGFS08}. For the case of LaFePO we find that
this actually makes the correspondence with the experimental data worse.  Much larger shifts are needed to bring the
predicted frequencies into agreement with experiment.

We finish by comparing the dHvA results with those of a recent angle-resolved photoemission spectroscopy study of the
same compound by Lu \emph{et al.} \cite{LuYMEACSHGFS08}.  The ARPES study could follow the band dispersions to
$\sim$0.5\,eV below the Fermi level.  They found that their band-dispersions were in agreement with the calculated
band-structure when renormalized by a factor 2 and shifted by $\sim$ 100meV.  This renormalisation is the same as we
found for quasiparticles \emph{at the Fermi level}, implying that the renormalisation occurs over the whole band,
rather than being localized to energies close to the Fermi level.  Renormalisation over a similar wide energy range was
also found in Sr$_2$RuO$_4$ \cite{Ingle05}.  The ARPES intensity close to the Fermi level gives evidence for small hole
pocket(s) at the zone centre and electron pocket(s) at the zone corner in agreement with our results and the
band-structure.  The ARPES results also show evidence for another large hole pocket cantered on $\Gamma$. Including
this pocket in the sum of electron and hole Fermi surface volumes leads to a very large electron imbalance of around
one electron per unit cell, suggesting a significant doping of the surface layer. This is possibly caused by the polar
nature of the cleaved surface \cite{LuYMEACSHGFS08}.

\section{Conclusions}

Measurements of the de Haas-van Alphen effect have been shown to provide a very precise probe of the Fermi surface
properties of LaFePO.  The Fermi surface topology deduced from the measurements is in good overall agreement with
density functional theory calculations, consisting of two hole sheets near the zone centre and two electron sheets
cantered on the zone corner. The data give an accurate picture of the warping of the electron sheets although there
remains some uncertainty about the correct identification for the hole sheets.   We hope these issues will be resolved
in the near future as larger, higher purity samples become available.

\section{Acknowledgements}
We thank E. A. Yelland, N. Fox, and M. F. Haddow for technical help and I.I.\ Mazin and O.K.\ Andersen for helpful
comments. This work was supported financially by EPSRC (U.K.) and the Royal Society. A.I.C. is grateful to the Royal
Society for financial support. Work at Stanford was supported by the U.S. DOE, Office of Basic Energy Sciences under
contract DE-AC02-76SF00515. Work performed at the NHMFL in Tallahassee, Florida, was supported by NSF Cooperative
Agreement No. DMR-0654118, by the State of Florida, and by the U.S. DOE.


\begin{thebibliography}{10}
\bibitem{angle}
The X-ray laue diffraction indicates that the exact axis of rotation between is
  $\sim 35^\circ$ (sample A) and $\sim 14^\circ$ (sample B) from the $a$ axis.
\bibitem{bulgenote}
We note that the `bulge' causes the maximum frequencies for bands 2 and 3 to
  have a weak angular dependence which would mean that their torque dHvA
  amplitudes would be small (see Eq.\ \ref{eqtorque}).
\end{thebibliography}
\end{document}